\documentclass[12pt]{article}

\usepackage{cite}
\usepackage{epsfig}
\usepackage{amsmath}
\usepackage{eufrak}
\usepackage{pstricks}

\newcommand{\mycaption}[1]{\caption{\sl #1}}

\makeatletter
\def\section{\@startsection {section}{1}{\z@}{+3.0ex plus +1ex minus
  +.2ex}{2.3ex plus .2ex}{\large\bf\boldmath}}
\def\subsection{\@startsection{subsection}{2}{\z@}{+2.5ex plus +1ex
minus +.2ex}{1.5ex plus .2ex}{\normalsize\bf\boldmath}}
\def\subsubsection{\@startsection{subsubsection}{3}{\z@}{+3.25ex plus
 +1ex minus +.2ex}{1.5ex plus .2ex}{\normalsize\it}}

\graphicspath{{.}{plots/}}

\begin{document}
\thispagestyle{empty}

\def\thefootnote{\fnsymbol{footnote}}

\begin{flushright}
\end{flushright}

\vspace{1cm}

\begin{center}

{\Large {\bf Three-loop vacuum integrals with arbitrary masses}}
\\[3.5em]
{\large
Ayres~Freitas
}

\vspace*{1cm}

{\sl
Pittsburgh Particle-physics Astro-physics \& Cosmology Center
(PITT-PACC),\\ Department of Physics \& Astronomy, University of Pittsburgh,
Pittsburgh, PA 15260, USA
}

\end{center}

\vspace*{2.5cm}

\begin{abstract}

Three-loop vacuum integrals are an important building block for the calculation
of a wide range of three-loop corrections. Until now, analytical results for
integrals with only one and two independent mass scales are known, but in the
electroweak Standard Model and many extensions thereof, one often encounters
more mass scales of comparable magnitude. For this reason, a numerical approach
for the evaluation of three-loop vacuum integrals with arbitrary mass pattern is
proposed here. Concretely, one can identify a basic set of three master integral
topologies. With the help of dispersion relations, each of these can be
transformed into one-dimensional or, for the most complicated case,
two-dimensional integrals in terms of elementary functions, which are suitable
for efficient numerical integration.

\end{abstract}

\setcounter{page}{0}
\setcounter{footnote}{0}

\newpage


\section{Introduction}

\noindent
The need for higher-order radiative corrections is growing more and more
important due to the increasing precision of measurements at the LHC and planned
future colliders. 
The anticipated precision of future experiments will require the evaluation of
three-loop corrections with arbitrary masses, $e.\,g.$ for precision electroweak
quantities (for a recent review see Ref.~\cite{review})
or a detailed understanding of the Higgs potential and its stability \cite{Martin:2014cxa}.
In this article, the calculation of general three-loop vacuum integrals is
considered, $i.\,e.$ integrals with vanishing external momentum and arbitrary
propagator masses.
Such integrals may arise in low-energy observables, in the coefficients of
low-momentum expansions (see $e.\,g.$ Ref.~\cite{2lvac}) or as building blocks
in more general three-loop calculations.

At the two-loop level, analytical formulae for general vacuum integrals have
been known for some time \cite{2lvac,2lvaca}. When expanding in powers of
$\epsilon = (4{-}D)/2$ within dimensional regularization, they can be written in
terms of polylogarithms. At the three-loop level, results for vacuum integrals
are only available for one
\cite{3lvaca,Broadhurst:1998rz,Chetyrkin:1999qi,sv}  and two
\cite{3lvac2m,Kalmykov:2006pu,Grigo:2012ji}\footnote{See also Ref.~\cite{3l2m} for some early
results on two-scale three-loop on-shell integrals, where several important
techniques for three-loop integrals were developed.} independent mass scales.
The derivation of analytical results for the class of two-scale three-loop
vacuum integrals requires the introduction of harmonic polylogarithms
\cite{Remiddi:1999ew}, and some cases are only known numerically
\cite{Grigo:2012ji}.

In light of these facts, a numerical approach to three-loop vacuum integrals with
general mass pattern appears most promising. In Ref.~\cite{gkp} a numerical
technique for the calculation of the four-propagator topology has been
presented.
In the following, a method for the evaluation of all relevant master integrals
is proposed, which is based
on dispersion relations. This technique has been previously
used for the numerical evaluation of two-loop self-energy and vertex integrals
\cite{disp2,disp2a}.
For the master integrals considered in this paper, the dispersion
relation approach leads to simple numerical integrals for their finite part.
For two of the three master integral topologies, one can obtain one-dimensional numerical
integral representations in terms of elementary functions. For the
most complicated case, the six-propagator master integral, one may construct a
two-dimensional integral in terms of elementary functions\footnote{Another
promising approach for the calculation of three-loop vacuum integrals is based
on the numerical integration of differential equations, see
Ref.~\cite{martin}.}.

Note that in some applications it may be necessary to evaluate the master
integrals to higher orders in $\epsilon$. This happens when a master integral is
multiplied by a coefficient that has poles in $1/\epsilon$. The 
method described in this paper, in its present form, is not suitable for such
situations.

\begin{figure}[t]
\begin{center}
\psset{linewidth=1pt}
\psset{dotsize=5pt}
\begin{tabular}{p{3cm}p{3cm}p{3cm}}
&& \\[-.2cm]
\begin{center}
\pscircle(0,0){1}%
\psdot(0,1)%
\psdot(0,-1)%
\psdot(-1,0)%
\psarc(-1,0){1.414}{-45}{45}%
\psarc(1,0){1.414}{135}{-135}%
\rput[r](-0.9,0.7){\small 1}%
\rput[r](-0.9,-0.7){\small 1}%
\rput[r](-0.5,0){\small 2}%
\rput[r](0.3,0){\small 3}%
\rput[l](1.05,0){\small 4}%
\end{center}
&
\begin{center}
\pscircle(0,0){1}%
\psdot(-0.707,0.707)%
\psdot(0,-1)%
\psdot(0.707,0.707)%
\psline(0,-1)(-0.707,0.707)%
\psline(0,-1)(0.707,0.707)%
\rput[r](-1,-0.4){\small 1}%
\rput[r](-0.45,-0.2){\small 2}%
\rput[l](1,-0.4){\small 4}%
\rput[l](0.45,-0.2){\small 3}%
\rput[t](0,0.9){\small 5}%
\end{center}
&
\begin{center}
\pscircle(0,0){1}%
\psdot(-0.866,0.5)%
\psdot(0,-1)%
\psdot(0.866,0.5)%
\psdot(0,0)%
\psline(0,0)(-0.866,0.5)%
\psline(0,0)(0.866,0.5)%
\psline(0,0)(0,-1)%
\rput[r](-1,-0.4){\small 1}%
\rput[r](-0.4,0){\small 2}%
\rput[l](1,-0.4){\small 4}%
\rput[l](0.4,0){\small 3}%
\rput[t](0,0.9){\small 5}%
\rput[l](0.1,-0.5){\small 6}%
\end{center}
\\[0.6cm]
\centering $U_4$ & \centering $U_5$ & \centering $U_6$
\end{tabular}
\end{center}
\vspace{-2.5ex}
\mycaption{Basic master integral topologies considered in this paper. The dot
indicates a propagator that is raised to the power 2.
\label{fig:diag1}}
\end{figure}

This article begins by defining the set of three-loop vacuum master integrals in
section~\ref{sc:def}. Each master integral topology is discussed in turn in
sections~\ref{sc:u4}--\ref{sc:u6}. Several special cases, which require a
modification of the integral representations, are treated separately in
sections~\ref{sc:u4} and~\ref{sc:u5}. For the most complicated master integral,
which is the subject of section~\ref{sc:u6}, no such special case has been
identified so far. The paper finishes with
some comments on the implementation of the numerical integrations in
section~\ref{sc:num} before concluding in section~\ref{sc:concl}.
Some useful formulae are collected in the appendix.


\section{Definition of basic integrals}
\label{sc:def}

\noindent
After trivial cancellations of numerator and denominator terms, a general scalar
three-loop vacuum integral may be written in the form
\begin{align}
M(\nu_1,\nu_2,\nu_3,\nu_4,\nu_5,\nu_6;\, &m_1^2,m_2^2,m_3^2,m_4^2,m_5^2,m_6^2)
\nonumber \\
= i\frac{e^{3\gamma_{\rm E}\epsilon}}{\pi^{3D/2}}
  & \int d^Dq_1\, d^Dq_2\, d^Dq_3 \;
  \frac{1}{[q_1^2-m_1^2]^{\nu_1} [(q_1-q_2)^2-m_2^2]^{\nu_2}} \nonumber \\
&\;\times
  \frac{1}{[(q_2-q_3)^2-m_3^2]^{\nu_3} [q_3^2-m_4^2]^{\nu_4}
	   [q_2^2-m_5^2]^{\nu_5} [(q_1-q_3)^2-m_6^2]^{\nu_6}}\,,
\end{align}
where $\epsilon = (4-D)/2$, $D$ is the number of dimensions in dimensional
regularization, and $\nu_i$ are integer numbers.
The complete set of three-loop vacuum integrals can be reduced to a small set of
master integrals with the help of integration-by-parts
identities~\cite{ibp}\footnote{The integration-by-parts technique may be
augmented by other algorithms \cite{weiglein} to increase the efficiency of the
reduction procedure.}. 
In most cases, not involving any special mass patterns, one can choose the
following basis of three master integrals, see Fig.~\ref{fig:diag1},
\begin{align}
M(2,1,1,1,0,0) &\equiv U_4(m_1^2,m_2^2,m_3^2,m_4^2), \\
M(1,1,1,1,1,0) &\equiv U_5(m_1^2,m_2^2,m_3^2,m_4^2,m_5^2), \\
M(1,1,1,1,1,1) &\equiv U_6(m_1^2,m_2^2,m_3^2,m_4^2,m_5^2,m_6^2), 
\end{align}
besides integrals that factorize into products of one- and two-loop
contributions. Two other simple integrals that are often encountered (see
Fig.~\ref{fig:diag2}) can be reduced to these three with the help of integration-by-parts
identities:
\begin{align}
M(1,1,1,1,0,0) &= 
 \frac{2}{3D-8}\bigl [
m_1^2\, U_4(m_1^2,m_2^2,m_3^2,m_4^2) + \text{cycl}_{1234}\bigr ], 
 \label{eq:m111100}
 \displaybreak[0] \\[1ex]
M(1,1,1,1,-1,0)\big|_{m_5=0} &= 
\biggl\{ \frac{2 m_1^2}{3(D-2)(3D-8)}
\bigl [ (D-2)m_1^2 + (7D-18)m_2^2 \nonumber \\
&\qquad - 2(D-3)(m_3^2+m_4^2) \bigr ] \,
 U_4(m_1^2,m_2^2,m_3^2,m_4^2) \nonumber \\
&\qquad+ \frac{1}{3}A_0(m_2^2)A_0(m_3^2)A_0(m_4^2) \biggr \} + 
 \Bigl\{ m_1 \leftrightarrow m_2 \Bigr\} \nonumber \\ &\quad +
 \Bigl\{ m_1 \leftrightarrow m_3, m_2 \leftrightarrow m_4 \Bigr\} +
 \Bigl\{ m_1 \leftrightarrow m_4, m_2 \leftrightarrow m_3 \Bigr\} \, ,
\end{align}
where ``cycl$_{1234}$'' refers to cyclic permutations of $\{m_1,m_2,m_3,m_4\}$, and
$A_0(m^2)$ is the standard one-loop vacuum function (see appendix~\ref{sc:12}).
\begin{figure}[tb]
\begin{center}
\psset{linewidth=1pt}
\psset{dotsize=5pt}
\begin{tabular}{p{3cm}p{3cm}}
& \\[-.2cm]
\begin{center}
\pscircle(0,0){1}%
\psdot(0,1)%
\psdot(0,-1)%
\psarc(-1,0){1.414}{-45}{45}%
\psarc(1,0){1.414}{135}{-135}%
\rput[r](-1.05,0){\small 1}%
\rput[r](-0.5,0){\small 2}%
\rput[r](0.3,0){\small 3}%
\rput[l](1.05,0){\small 4}%
\end{center}
&
\begin{center}
\pscircle(0,0){1}%
\psdot(-0.707,0.707)%
\psdot(0,-1)%
\psdot(0.707,0.707)%
\psdot[dotsize=8pt,dotstyle=x](0,0.97)%
\psline(0,-1)(-0.707,0.707)%
\psline(0,-1)(0.707,0.707)%
\rput[r](-1,-0.4){\small 1}%
\rput[r](-0.45,-0.2){\small 2}%
\rput[l](1,-0.4){\small 4}%
\rput[l](0.45,-0.2){\small 3}%
\rput[t](0,0.8){\small 5}%
\end{center}
\\[0.8cm]
\centering $M(1,1,1,1,0,0)$ & \centering M(1,1,1,1,-1,0) \newline ($m_5$=0)
\end{tabular}
\end{center}
\vspace{-2ex}
\mycaption{Other basic scalar integrals, which can be expressed in
terms of the
ones in Fig.~\ref{fig:diag1}. The cross
indicates a propagator that is raised to the power $-1$.
\label{fig:diag2}}
\end{figure}

The master integrals $U_4$, $U_5$ and $U_6$ have the following symmetry
properties:
\begin{itemize}
\item
$U_4(m_1^2,m_2^2,m_3^2,m_4^2)$ is symmetric under arbitrary permutations of
$m_{2,3,4}$.
\item
$U_5(m_1^2,m_2^2,m_3^2,m_4^2,m_5^2)$ is symmetric under the replacements $\{m_1
\leftrightarrow m_2\}$, $\{m_3 \leftrightarrow m_4\}$, and 
$\{m_1 \leftrightarrow m_3, \, m_2 \leftrightarrow
m_4\}$, as well as any combination thereof.
\item
$U_6(m_1^2,m_2^2,m_3^2,m_4^2,m_5^2,m_6^2)$ is symmetric under the replacements
$\{m_2 \leftrightarrow m_3,\, m_1 \leftrightarrow m_4\}$,
$\{m_2 \leftrightarrow m_6,\, m_4 \leftrightarrow m_5\}$,
$\{m_1 \leftrightarrow m_6,\, m_3 \leftrightarrow m_5\}$,
and any combination thereof.
\end{itemize}


\section{\boldmath $U_4$}
\label{sc:u4}

\subsection{General case}
\label{sc:u4gen}

\noindent
Let us begin with a dispersion relation for the double bubble loop integral in
Fig.~\ref{fig:diag3},
\begin{align}
I_{\rm db}(p^2,m_1^2,m_2^2,m_3^2,m_4^2) &\equiv B_{0,m_1}(p^2,m_1^2,m_2^2)
B_0(p^2,m_3^2,m_4^2)= \int_0^\infty ds \, \frac{\Delta I_{\rm db}(s)}{s-p^2-i\varepsilon}, 
\label{eq:disp1} \\[1ex]
\Delta I_{\rm db}(s,m_1^2,m_2^2,m_3^2,m_4^2) &= 
 \Delta B_{0,m_1}(s,m_1^2,m_2^2) \, B_0(s,m_3^2,m_4^2) \nonumber \\
 &\quad +
 B_{0,m_1}(s,m_1^2,m_2^2) \, \Delta B_0(s,m_3^2,m_4^2),
\end{align}
where $B_0$ is the standard scalar one-loop self-energy function (see appendix~\ref{sc:12}), and $B_{0,m_1}
(p^2,m_1^2,m_2^2) = \frac{\partial}{\partial m_1^2}B_0(p^2,m_1^2,m_2^2)$. The
discontinuities of these two functions are denoted by $\Delta B_0$ and $\Delta
B_{0,m_1}$, respectively. In  $D=4$ dimensions, they are given by
\begin{align}
\Delta B_0(s,m_a^2,m_b^2) &= \frac{1}{s}\lambda(s,m_a^2,m_b^2) \,
\Theta\bigl(s-(m_a+m_b)^2\bigr)\,, \\
\Delta B_{0,m_1}(s,m_a^2,m_b^2) &= \frac{m_a^2-m_b^2-s}{s\,\lambda(s,m_a^2,m_b^2)} \,
\Theta\bigl(s-(m_a+m_b)^2\bigr)\,,
\intertext{where}
\lambda(x,y,z) &= \sqrt{x^2+y^2+z^2-2(xy+yz+xz)}\,, \label{eq:lambda}
\end{align}
and $\Theta(t)$ is the Heaviside step function.
\begin{figure}[tb]
\begin{center}
\vspace{1cm}
\psset{linewidth=1pt}
\psset{dotsize=5pt}
\pscircle(-0.6,0){0.6}%
\pscircle(0.6,0){0.6}%
\psdot(0,0)%
\psdot(-1.2,0)%
\psdot(1.2,0)%
\psdot(-0.6,-0.6)%
\psline(-2,0)(-1.2,0)%
\psline(2,0)(1.2,0)%
\rput[t](-0.9,-0.7){\small 1}%
\rput[t](-0.3,-0.7){\small 1}%
\rput[b](-0.6,0.7){\small 2}%
\rput[t](0.6,-0.7){\small 3}%
\rput[b](0.6,0.7){\small 4}%
\rput[b](-1.6,0.1){$\stackrel{{\displaystyle p}}{\rightarrow}$}
\end{center}
\vspace{2ex}
\mycaption{Double-bubble two-loop sub-topology.
\label{fig:diag3}}
\end{figure}

Inserting the dispersion relation eq.~\eqref{eq:disp1} into the three-loop
integral $U_4$, one obtains
\begin{align}
U_4(m_1^2,m_2^2,m_3^2,m_4^2) &= -\frac{e^{\gamma_{\rm E}\epsilon}}{i\pi^{D/2}} 
 \int d^Dq_3 \int_0^\infty ds \, \frac{\Delta I_{\rm db}(s)}{q_3^2-s+i\varepsilon} \\
&= -\int_0^\infty ds \, A_0(s) \, \Delta I_{\rm db}(s)\,. \label{eq:disp2}
\end{align}
This integral is divergent, and thus a numerical integration of
\eqref{eq:disp2} in $D=4$ dimension is not possible. Instead one may consider
the sum
\begin{align}
U_4(m_1^2,m_2^2,m_3^2,m_4^2) &= U_4(m_1^2,m_2^2,0,0)+
U_4(m_1^2,0,m_3^2,0) + U_4(m_1^2,0,0,m_4^2) \nonumber \\ 
&\quad- 2\, U_4(m_1^2,0,0,0) +
U_{4,\rm sub}(m_1^2,m_2^2,m_3^2,m_4^2)\,.
\label{eq:u4red}
\end{align}
The $U_4$ master integrals with one or two non-zero masses can be calculated
analytically, with results collected in appendix~\ref{sc:ana}. The first four
terms on the right-hand side of eq.~\eqref{eq:u4red} precisely reproduce the
divergencies of the general $U_4$ integral on the left-hand side, such that
the remainder
$U_{4,\rm sub}$ is finite and can be integrated numerically. It is given by
\begin{align}
&U_{4,\rm sub}(m_1^2,m_2^2,m_3^2,m_4^2) 
 = -\int_0^\infty ds \, A_{0,\rm fin}(s) \, \Delta I_{\rm db,sub}(s)\,, 
 \label{eq:u4sub} \\[1ex]
&\Delta I_{\rm db,sub}(s,m_1^2,m_2^2,m_3^2,m_4^2) = \nonumber \\
&\qquad \Delta B_{0,m_1}(s,m_1^2,m_2^2) \; \text{Re}\bigl\{ B_0(s,m_3^2,m_4^2)
- B_0(s,0,0) \bigr\} \nonumber \\
 &\qquad - \Delta B_{0,m_1}(s,m_1^2,0) \; \text{Re} \bigl\{ B_0(s,0,m_3^2) + B_0(s,0,m_4^2)
- 2 B_0(s,0,0) \bigr\} \nonumber \\
&\qquad + \text{Re}\bigl\{B_{0,m_1}(s,m_1^2,m_2^2)\bigr\} \, \bigl [ \Delta B_0(s,m_3^2,m_4^2) -
 \Delta B_0(s,0,0) \bigr ] \nonumber \\
&\qquad - \text{Re} \bigl\{B_{0,m_1}(s,m_1^2,0)\bigr\} \, \bigl [ \Delta B_0(s,0,m_3^2) + 
 \Delta B_0(s,0,m_4^2) - 2\, \Delta B_0(s,0,0) \bigr ]\,,
\end{align}
which has been written in a way that makes its finiteness manifest. The
divergent part of the $A_0$ function in eq.~\eqref{eq:u4sub} integrates to zero
and thus can be ignored.

\subsection{Special case: \boldmath $m_1=0$}

\noindent
A special treatment is required for the case $m_1=0$, when
$U_4(m_1^2,m_2^2,m_3^2,m_4^2)$ develops an infrared divergence. This singularity
occurs in the limit when the first loop momentum goes on-shell, $q_1^2 \to 0$.
In this limit $U_4(0,m_2^2,m_3^2,m_4^2)$ factorizes into the product
$B_0(0,0,0) \,T_3(m_2^2,m_3^2,m_4^2)$. Note that $B_0(0,0,0)=0$ in dimensional
regularization, but it has been kept here for illustration. By alternatively
considering a mass regulator for the infrared divergence, one arrives
at the identity
\begin{align}
U_4(0,m_2^2,m_3^2,m_4^2) &= B_0(0,0,0) \,T_3(m_2^2,m_3^2,m_4^2) -
B_0(0,\delta^2,\delta^2) \,T_3(m_2^2,m_3^2,m_4^2) \nonumber \\
&\quad+ U_4(\delta^2,m_2^2,m_3^2,m_4^2)
+ {\cal O}(\delta^2), \label{eq:u40}
\end{align}
where $T_3$ is the basic scalar two-loop vacuum integral (see
appendix~\ref{sc:12}), and $\delta$ is an infinitesimal mass parameter. In
principle, one could directly evaluate $U_4(\delta^2,m_2^2,m_3^2,m_4^2)$ for a
small numerical value of $\delta$ according to the prescription in the previous
subsection, although one has to pay the price of having numerical cancellations
between the two $\log \delta$ terms from the last two terms in eq.~\eqref{eq:u40}.

Alternatively, it is possible to extract the $\log \delta$ dependence explicitly
from $U_4(\delta^2,m_2^2,m_3^2,m_4^2)$. For this purpose, let us consider the
following small $\delta$ expansions:
\begin{align}
\text{Re} \bigl\{ B_{0,m_1}(s,\delta^2,m_2^2)\bigr\} &=
\frac{1}{s-m_2^2}\biggl [\biggl (1+\frac{m_2^2}{s} \biggr ) \,\text{Re}\Bigl\{\log
\frac{m_2^2-s}{m_2^2}\Bigr \} - \log \frac{\delta^2}{m_2^2} \biggr ]
\nonumber \\
& \quad  - \pi^2 \delta(s-m_2^2) + {\cal O}(\delta^2), \\
\text{Re} \bigl\{ B_{0,m_1}(s,\delta^2,0)\bigr\} &=
\frac{1}{s} \log \frac{s}{\delta^2} + {\cal O}(\delta^2), \\
\int_0^\infty ds\, \Delta B_{0,m_1}(s,\delta^2,m_2^2)\, f(s) &= 
 \int_0^\infty ds\, \Delta B_{0,m_1}(s,0,m_2^2)\,\biggl [
  f(s) - f(m_2^2)\frac{m_2^2}{s} \biggr ] \nonumber \\ &\quad + 
\biggl (1+ \log
 \frac{\delta^2}{m_2^2}\biggr ) f(m_2^2) + {\cal O}(\delta^2),
\end{align}
where $f(s)$ is some arbitrary well-behaved function that does not depend on $\delta$.
For the remaining terms in the integrand, involving $B_0$ and $\Delta B_0$ 
functions, one can simply set $\delta$ to zero. One then obtains
\begin{align}
&U_{4,\rm sub}(\delta^2,m_2^2,m_3^2,m_4^2) 
 = -\int_0^\infty ds \, A_{0,\rm fin}(s) \, \Delta I_{\rm db,sub,0}(s) 
 + U_{4,\rm add,0}(\delta^2,m_2^2,m_3^2,m_4^2)\,, \displaybreak[0] \\[1ex]
&\Delta I_{\rm db,sub,0}(s,m_2^2,m_3^2,m_4^2) = \nonumber \\
&\qquad \Delta B_{0,m_1}(s,0,m_2^2) \biggl [
\begin{aligned}[t] &\text{Re}\bigl\{ B_0(s,m_3^2,m_4^2)
- B_0(s,0,0) \bigr\} \\ &- \frac{m_2^2\, A_{0,\rm fin}(m_2^2)}{s\, A_{0,\rm fin}(s)}\,
 \text{Re}\bigl\{ B_0(m_2^2,m_3^2,m_4^2)
- B_0(m_2^2,0,0) \bigr\} \biggr ] \end{aligned}
\nonumber \\
 &\qquad - \Delta B_{0,m_1}(s,0,0) \; \text{Re} \bigl\{ B_0(s,0,m_3^2) + B_0(s,0,m_4^2)
- 2 B_0(s,0,0) \bigr\} \nonumber \\
&\qquad + \frac{s+m_2^2}{s(s-m_2^2)} \text{Re}\Bigl\{\log
\frac{m_2^2-s}{m_2^2}\Bigr \} \, \bigl [ \Delta B_0(s,m_3^2,m_4^2) -
 \Delta B_0(s,0,0) \bigr ] \nonumber \\
&\qquad - \frac{1}{s} \log \frac{s}{m_2^2} \, \bigl [ \Delta B_0(s,0,m_3^2) + 
 \Delta B_0(s,0,m_4^2) - 2\, \Delta B_0(s,0,0) \bigr ]\,, 
\displaybreak[0] \\[1ex]
&U_{4,\rm add,0}(\delta^2,m_2^2,m_3^2,m_4^2) = \nonumber \\
&\qquad -\biggl (1+ \log \frac{\delta^2}{m_2^2}\biggr ) \, A_{0,\rm fin}(m_2^2)\,
 \text{Re}\bigl\{ B_0(m_2^2,m_3^2,m_4^2)
- B_0(m_2^2,0,0) \bigr\} \nonumber \\
&\qquad +\log \frac{\delta^2}{m_2^2} \int_0^\infty ds \,
 \frac{1}{s-m_2^2}A_{0,\rm fin}(s)\,\bigl [ \Delta B_0(s,m_3^2,m_4^2) -
 \Delta B_0(s,0,0) \bigr ] \nonumber \\
&\qquad -\log \frac{\delta^2}{m_2^2} \int_0^\infty ds \,
 \frac{1}{s}A_{0,\rm fin}(s)\,\bigl [ \Delta B_0(s,0,m_3^2) + 
 \Delta B_0(s,0,m_4^2) - 2\, \Delta B_0(s,0,0) \bigr ] \nonumber \\
&\qquad +\pi^2 A_{0,\rm fin}(m_2^2)\, \bigl [ \Delta B_0(m_2^2,m_3^2,m_4^2) -
 \Delta B_0(m_2^2,0,0) \bigr ] \displaybreak[0] \\
&\quad = - \log \frac{\delta^2}{m_2^2} \Bigl [
T_3(m_2^2,m_3^2,m_4^2) - \sum_{i=2}^4 T_3(m_i^2,0,0) \Bigr ] \nonumber \\
&\qquad \begin{aligned}[b] -A_{0,\rm fin}(m_2^2)\, \Bigl [
 &\text{Re}\bigl\{ B_0(m_2^2,m_3^2,m_4^2)
- B_0(m_2^2,0,0)  \bigr\} \\
&- \pi^2\, \Delta B_0(m_2^2,m_3^2,m_4^2) + \pi^2 \,
\Delta B_0(m_2^2,0,0) \Bigr ]\,.
\end{aligned}
\end{align}


\section{\boldmath $U_5$}
\label{sc:u5}

\subsection{General case}
\label{sc:u5gen}

\noindent
The master integral $U_5$ can be addressed with a dispersion
relation similar to eq.~\ref{eq:disp1}. As in the previous section, one first
needs to subtract the divergencies to arrive at a finite integral suitable for
numerical evaluation. For this purpose, it is useful to consider 
the following relation, which has been derived from
integration-by-parts identities:
\begin{multline}
U_5(m_1^2,m_2^2,m_3^2,m_4^2,m_5^2) = F\bigl [A_0(m_i),\,
T_3(m_i,m_j,m_k),\, U_4(m_i,m_j,m_k,m_l) \bigr ] \\
+ \frac{\lambda^2_{125}
\lambda^2_{345}}{(3-D)^2(m_2^2-m_1^2+m_5^2)(m_3^2-m_4^2+m_5^2)}
 M(2,1,1,2,1,0)\,, \label{eq:u5red}
\end{multline}
where $F[...]$ is a linear combination of $U_4$ functions and products of $A_0$
and $T_3$ functions,
whose explicit form is too lengthy to be included here. It is provided as an
ancillary file in {\sc Mathematica} format in the arXiv submission. Furthermore,
$
\lambda_{ijk} = \lambda(m_i^2,m_j^2,m_k^2).
$

For $m_{1,4} > 0$, $M(2,1,1,2,1,0)$ is finite and can be computed numerically.
If $m_1=0$ ($m_4=0$) one can make the trivial replacement $m_1 \leftrightarrow
m_2$ ($m_3 \leftrightarrow m_4$). The case when both masses in a sub-loop bubble
are zero ($e.\,g.$ $m_1=m_2=0$) will be treated in section~\ref{sc:u50}.

Following the approach of section~\ref{sc:u4gen}, $M(2,1,1,2,1,0)$ can be
expressed in terms of a dispersion integral. The relevant dispersion relation
reads
\begin{align}
I_{\rm db2}(p^2,m_1^2,m_2^2,m_3^2,m_4^2) &\equiv B_{0,m_1}(p^2,m_1^2,m_2^2)
B_{0,m_1}(p^2,m_4^2,m_3^2)= \int_0^\infty ds \, \frac{\Delta I_{\rm db2}(s)}{s-p^2-i\varepsilon}, 
 \\[1ex]
\Delta I_{\rm db2}(s,m_1^2,m_2^2,m_3^2,m_4^2) &= 
 \Delta B_{0,m_1}(s,m_1^2,m_2^2) \, B_{0,m_1}(s,m_4^2,m_3^2) \nonumber \\
 &\quad +
 B_{0,m_1}(s,m_1^2,m_2^2) \, \Delta B_{0,m_1}(s,m_4^2,m_3^2).
\end{align}
Inserting this expression into the $M(2,1,1,2,1,0)$ integral, one finds
\begin{align}
M(2,1,1,2,1,0;\,m_1^2,m_2^2,m_3^2,m_4^2,m_5^2) &= -\frac{e^{\gamma_{\rm E}\epsilon}}{i\pi^{D/2}} 
 \int d^Dq_3 \int_0^\infty ds \, \frac{\Delta I_{\rm db2}(s)}{%
 [q_3^2-s+i\varepsilon][q_3^2-m_5^2+i\varepsilon]} \nonumber \\
&= -\int_0^\infty ds \, B_0(0,s,m_5^2) \, \text{Re}\bigl\{ \Delta I_{\rm db2}(s)
\bigr\}\,.
\label{eq:dispu5}
\end{align}
The divergent part of the $B_0$ function in eq.~\eqref{eq:dispu5} integrates to
zero, and thus one can immediately replace $B_0(...) \to B_{0,\rm fin}(\dots)$.

By combining eq.~\eqref{eq:dispu5} with eq.~\eqref{eq:u5red} and evaluating
the $U_4$ functions in eq.~\eqref{eq:u5red} according to the procedure discussed in
the previous section, one thus arrives at a final result for $U_5$.

\subsection{Special case: \boldmath $m_1^2=m_2^2+m_5^2$}

\noindent
The formula \eqref{eq:u5red} is not valid for the case
$m_1^2=m_2^2+m_5^2$.\footnote{The
similar case $m_4^2=m_3^2+m_5^2$ can be mapped to the former by flipping the two
sub-bubbles of $U_5$, $i.\,e.$ $\{m_1 \leftrightarrow m_3, \, m_2
\leftrightarrow m_4\}$.} As long as $m_5>0$, this problem can be avoided by
simply flipping $m_1 \leftrightarrow m_2$. A special treatment is needed,
however, for the case $m_5=0$ and $m_1=m_2$. In this case, instead of eq.~\eqref{eq:u5red}
one can use the modified formula
\begin{align}
&U_5(m_1^2,m_1^2,m_3^2,m_4^2,0) = F'\bigl [A_0(m_i),\,
T_3(m_i,m_j,m_k),\, U_4(m_i,m_j,m_k,m_l) \bigr ] \nonumber \\
&\quad+ \frac{4 m_1^2 m_4^2 (m_4^2-m_3^2)\, \lambda^2(4 m_1^2,m_3^2,m_4^2)}%
{(3-D)[(3-D)m_4^2\,\lambda^2(4
m_1^2,m_3^2,m_4^2)+2m_1^2(4m_1^2m_3^2+4m_1^2m_4^2-8m_3^2m_4^2-m_3^4-m_4^4)]} 
\nonumber \\
&\qquad\times
 M(2,1,1,2,1,0)\,. 
\end{align}
The full expression for the terms involving $A_0$, $T_3$ and $U_4$ functions is
again provided in the ancillary file.
The finite remainder $M(2,1,1,2,1,0)$ can be evaluated as above.

\subsection{Special case: \boldmath $m_1=m_2=0$}
\label{sc:u50}

\noindent
While the integral $U_5$ is infrared finite for $m_1=m_2=0$, the application of the
procedure in section \ref{sc:u5gen} would lead to spurious divergencies in
several terms in eq.~\eqref{eq:u5red} that only cancel in the sum. Instead, the
following integration-by-parts relation proves more useful for this situation:
\begin{multline}
U_5(0,0,m_3^2,m_4^2,m_5^2) = F''\bigl [A_0(m_i),\,
T_3(m_i,m_j,m_k),\, U_4(m_i,m_j,m_k,m_l) \bigr ] \\
\begin{aligned}[b]
+\, \frac{\lambda^2_{345}}{(3-D)(m_3^2-m_4^2+m_5^2)} \biggl [
 &\frac{m_5^2-m_4^2}{3-D}\,
 M(2,1,1,2,1,0;\,m_4^2,0,m_3^2,m_4^2,m_5^2)\\
 &-m_4^2 \,M(1',1,1,2,1,0;\,m_4^2,0,m_3^2,m_4^2,m_5^2) \,, 
 \end{aligned} \label{eq:u50red}
\end{multline}
where 
\begin{align}
&M(1',1,1,2,1,0;\,m_4^2,0,m_3^2,m_4^2,m_5^2) \nonumber \\
&\qquad \begin{aligned}[b]
\equiv  i\frac{e^{3\gamma_{\rm E}\epsilon}}{\pi^{3D/2}}
  \int d^Dq_1\, d^Dq_2\, d^Dq_3 \;
  &\frac{1}{q_1^2[q_1^2-m_4^2] (q_1-q_2)^2} \\
 &\times \frac{1}{[(q_2-q_3)^2-m_3^2] [q_3^2-m_4^2]^2
	   [q_2^2-m_5^2]}\,,
\end{aligned}
\displaybreak[0] \\
&\qquad = -\int_0^\infty ds \, B_0(0,s,m_5^2) \, \text{Re}\bigl\{ \Delta I_{\rm
db3}(s) \bigr \}\,,
\intertext{with}
&\begin{aligned}[b]\Delta I_{\rm db3}(s,m_4^2,0,m_3^2,m_4^2) = 
 {} &\frac{1}{m_4^2}\bigl [ \Delta B_0(s,0,m_4^2)-\Delta B_0(s,0,0) \bigr] \, B_{0,m_1}(s,m_4^2,m_3^2) \nonumber \\
 & +
 \frac{1}{m_4^2}\bigl [ B_0(s,0,m_4^2)-B_0(s,0,0) \bigr] \, \Delta
  B_{0,m_1}(s,m_4^2,m_3^2)\,.
\end{aligned}
\end{align}
$M(1',1,1,2,1,0)$ is finite (for $m_4>0$) and can be evaluated numerically.
As above, the full expression $F''$ involving $A_0$, $T_3$ and $U_4$ functions is
provided in the ancillary file.

The condition $m_4>0$ cannot be fulfilled, even with the help of the symmetry
relations at the end of section~\ref{sc:def}, only if $m_1=m_2=m_3=m_4=0$, but this
case can be solved analytically and is given in the appendix.


\section{\boldmath $U_6$}
\label{sc:u6}

\begin{figure}[tb]
\rule{0mm}{0mm}\\[-.2cm]
\begin{center}
\psset{linewidth=1pt}
\psset{dotsize=5pt}
\pscircle(0,0){1}%
\psdot(-1,0)%
\psdot(0,-1)%
\psdot(1,0)%
\psdot(0,1)%
\psline(-1.5,0)(-1,0)%
\psline(1,0)(1.5,0)%
\psline(0,1)(0,-1)%
\rput[r](-1,-0.7){\small 1}%
\rput[r](-1,0.7){\small 2}%
\rput[l](1,-0.7){\small 4}%
\rput[l](1,0.7){\small 3}%
\rput[l](0.15,0){\small 6}%
\rput[rb](-1.8,-0.1){$\stackrel{{\displaystyle p}}{\rightarrow}$}%
\end{center}
\vspace{3ex}
\mycaption{Two-loop five-propagator self-energy master integral, $T_5$. 
\label{fig:2lmaster}}
\end{figure}
\noindent
Without loss of generality, it is assumed in the following that $m_6 \geq m_i$
for $i=1,\dots,5$. Following Ref.~\cite{Broadhurst:1998rz}, one can write the
$U_6$ master integral in terms of
\begin{align}
&U_6(m_1^2,m_2^2,m_3^2,m_4^2,m_5^2,m_6^2) =
 -\int_0^\infty  ds \, B_0(0,s,m_5^2) \, \text{Re}\bigl\{\Delta
 T_5(s,m_1^2,m_2^2,m_3^2,m_4^2,m_6^2)\bigr\}, \label{eq:u6a}
\end{align}
where $\Delta T_5$ is the discontinuity of the two-loop five-propagator
self-energy master integral shown in Fig.~\ref{fig:2lmaster}. Evidently,
eq.~\eqref{eq:u6a} is divergent, but one can arrive at a finite integral by
considering the difference between the general $U_6$ integral and a simpler
$U_6$ with only one independent mass scale:
\begin{align}
&U_6(m_1^2,m_2^2,m_3^2,m_4^2,m_5^2,m_6^2) =
U_6(m_6^2,m_6^2,m_6^2,m_6^2,m_6^2,m_6^2)  \nonumber \\
&\qquad\qquad\qquad \begin{aligned}[b]
- \int_0^\infty  ds \, \Bigl [
 &B_{0,\rm fin}(0,s,m_5^2) \, 
 \text{Re}\bigl\{\Delta T_5(s,m_1^2,m_2^2,m_3^2,m_4^2,m_6^2) \bigr\}\\ -
 &B_{0,\rm fin}(0,s,m_6^2) \, \text{Re}\bigl\{\Delta
 T_5(s,m_6^2,m_6^2,m_6^2,m_6^2,m_6^2)\bigr\} \Bigr ] \end{aligned}
 \label{eq:u6disp}
\end{align}
The special case of $U_6$ with identical masses in all six propagators is known
analytically \cite{Broadhurst:1998rz},
\begin{align}
U_6(m_6^2,m_6^2,m_6^2,m_6^2,m_6^2,m_6^2) &= (m_6^2)^{-3\epsilon}\frac{2\zeta(3)}{\epsilon}
+ 6\zeta(3) - 17 \zeta(4) - 4 \zeta(2) \log^2 2 + \tfrac{2}{3}\log^4 2 
\nonumber \\ &\quad + 16\,
\text{Li}_4 (\tfrac{1}{2}) -4 \,\text{Cl}_2^2(\pi/3),
\end{align}
where $\zeta(x)$ is the Riemann zeta function, and Cl$_2$ is the second Clausen
function. The discontinuity $\Delta T_5$ can be written as a sum over the
possible two- and three-particle cuts \cite{Broadhurst:1987ei,disp2a},
\begin{align}
&\Delta T_5(s,m_1^2,m_2^2,m_3^2,m_4^2,m_6^2) =
\Delta T_5^{\rm (2a)} + 
\Delta T_5^{\rm (2b)} + 
\Delta T_5^{\rm (3a)} + 
\Delta T_5^{\rm (3b)}, \displaybreak[0] \\
&\Delta T_5^{\rm (2a)} = -\Delta B_0(s,m_1^2,m_2^2) \int_{(m_3+m_4)^2}^\infty dt \,
 \frac{[\Delta C_0(t-i\varepsilon,m_1^2,m_2^2; m_3^2,m_4^2,m_6^2)]^*}{t-s+i\varepsilon}\,,
\label{eq:t2a} \displaybreak[0] \\
&\Delta T_5^{\rm (2b)} = [\Delta T_5^{\rm (2a)}(m_2 \leftrightarrow m_3, 
m_1 \leftrightarrow m_4)]^*, \displaybreak[0] \\
&\Delta T_5^{\rm (3a)} = \int_{(m_4+m_6)^2}^{(\sqrt{s}-m_2)^2} dt \,
\frac{\Delta B_0(s,t,m_2^2) 
 \,\Delta C_0(t,m_2^2,s; m_4^2,m_6^2,m_3^2)}{t-m_1^2+i\varepsilon}\,,
\label{eq:t3a} \displaybreak[0] \\
&\Delta T_5^{\rm (3b)} = \Delta T_5^{\rm (3a)}(m_2 \leftrightarrow m_3, 
m_1 \leftrightarrow m_4), 
\end{align}
where
\begin{align}
\Delta C_0(s,p_2^2,p_3^2; m_1^2,m_2^2,m_3^2) &=
- \frac{2\, \text{Artanh}(b/a)}{\lambda(s,p_2^2,p_3^2)} \, 
 \Theta\big(s-(m_1+m_2)^2\bigr)\,, \\[.5ex]
a(s,p_2^2,p_3^2; m_1^2,m_2^2,m_3^2) &=
 s(s+2m_3^2-p_2^2-p_3^2-m_1^2-m_2^2) + (p_3^2-p_2^2)(m_1^2-m_2^2)\,, \\[.5ex]
b(s,p_2^2,p_3^2; m_1^2,m_2^2) &= \lambda(s,p_2^2,p_3^2)\,
\lambda(s,m_1^2,m_2^2)\,. \label{eq:delc0}
\end{align}
In general, eq.~\eqref{eq:t2a} also contains a contribution from an anomalous
threshold \cite{disp2a}, but for the mass ordering $m_6 \geq m_i$ it is never
encountered.

Combining eqs.~\eqref{eq:u6disp}--\eqref{eq:delc0} one arrives at a
two-dimensional integral representation for $U_6$ in terms of elementary
functions, such a logarithms and square roots.

Alternatively, the two-particle cut contributions (2a) and (2b) can also be
written directly in terms of the tree-point function $C_0$,
\begin{align}
&\Delta T_5^{\rm (2a)} = -\Delta B_0(s,m_1^2,m_2^2) \,
[C_0(s,m_1^2,m_2^2; m_3^2,m_4^2,m_6^2)]^*,
\label{eq:t2aC0} \displaybreak[0] \\
&\Delta T_5^{\rm (2b)} = [\Delta T_5^{\rm (2a)}(m_2 \leftrightarrow m_3, 
m_1 \leftrightarrow m_4)]^*.
\end{align}
Explicit formulae for the $C_0$ function in terms of polylogarithms can be found
in Ref.~\cite{c0}. Therefore, by using eq.~\eqref{eq:t2aC0} one arrives at
one-dimensional numerical integrals for the (2a) and (2b) terms, but at the cost
of requiring non-elementary functions in the integrand.

The numerical dispersion relation based on eq.~\eqref{eq:u6disp} proved to be
applicable to a large variety of different mass configurations, without the need
to consider special cases as in the previous two sections.


\section{Numerical integration and checks}
\label{sc:num}

\noindent
The dispersion relation techniques introduced in the previous sections lead to
one- and two-dimensional integrals, which can be efficiently evaluated with
numerical methods. In many cases, one encounters integrals of the form
\begin{align}
\int_{s_0}^\infty ds \, \frac{f(s)}{s-s'\pm i \varepsilon}\,.
\end{align}
For $s' > s_0$, these can be split into a residuum contribution and principal
value integral, resulting in
\begin{align}
\int_{s_0}^\infty ds \, \frac{f(s)}{s-s'\pm i \varepsilon} =
\mp i\pi f(s') + \int_{s_0}^{s'} ds \, \frac{f(s)-f(2s'-s)}{s-s'} +
\int_{2s'-s_0}^\infty ds \, \frac{f(s)}{s-s'}\,.
\end{align}
In this expression, the integrand of the first remaining integral is now regular
at the point $s=s'$. If $f(s)$ is real, the residuum contribution can be dropped
since only the real part is needed for the evaluation of the three-loop vacuum
integrals.

Results have been tested both within {\sc Mathematica 10} \cite{mathematica} and
in C++ using the Gauss-Kronrod algorithm from the {\sc Quadpack} library
\cite{quadpack}. The following checks were carried out for the $U_4$ and $U_5$
master integrals:  At least ten digits agreement were obtained when comparing
with the numbers in Ref.~\cite{Broadhurst:1998rz}, the integrals eq.~(24) of
Ref.~\cite{Chetyrkin:1999qi}, and the results for the $J_{\rm 7a}^{(3)}$,
$J_{\rm 7b}^{(3)}$, $J_{\rm 9a}^{(3)}$, $J_{\rm 9b}^{(3)}$, $J_{\rm 10a}^{(3)}$
and $J_{\rm 10b}^{(3)}$ integrals in Ref.~\cite{Grigo:2012ji}\footnote{Note that
some of the results in Ref.~\cite{Grigo:2012ji} have been obtained through 
numerical evaluation and interpolation, which may limit the level of achievable
numerical agreement.}. These correspond to $U_4$, $U_5$ and $U_6$ integrals with
one or two different mass scales. Similarly good agreement was obtained for the
$J_1^{(3)}$, $J_3^{(3)}$, $J_{\rm 8a}^{(3)}$, $J_{\rm 8b}^{(3)}$ integrals in
Ref.~\cite{Grigo:2012ji}, which are not among the set of master integrals in
Fig.~\ref{fig:diag1}, but which can easily be derived using
eq.~\eqref{eq:m111100}.

A public computer code, currently under development, will be presented in a
future publication.


\section{Conclusions}
\label{sc:concl}

\noindent
A general three-loop vacuum integral can be reduced to one of the three master
integral topologies shown in Fig.~\ref{fig:diag1}. The master integrals can be
evaluated analytically for all cases with one independent mass scale and for
several known cases with two independent mass scales. For general mass patterns,
however, numerical integration techniques need to be employed.

Using a method based on dispersion relations, the four-propagator master
integral $U_4$ and the five-propagator master integral $U_5$ can be expressed as
one-dimensional numerical integrals in terms of elementary functions, such a
logarithms and square roots. Similarly, the six-propagator master $U_6$ can be
representated by a two-dimensional numerical integral in terms of elementary
functions. These numerical integrals can be efficiently evaluated numerically,
yielding results with at least ten digits precision for the $U_4$ and $U_5$
functions and eights digit precision for the $U_6$ function.

To ensure that the numerical integrals are UV-finite, the divergent pieces of
the $U_4$, $U_5$ and $U_6$ must be subtracted beforehand. This can be achieved
by subtracting suitable linear combinations of special cases of these integrals,
which can be evaluated analytically. A public computer code that carries out
these subtractions and the numerical integrations is in development and will be
presented in a future publication.

The technique described in this article has been tailored for the evaluation of
the UV-divergent and finite parts of the master integrals. In principle, the
same dispersion relation approach could be used for the calculation of higher
order terms in $\epsilon$, but its concrete realization would require a
significant amount of additional work.


\section*{Acknowledgments}

\noindent
It is a pleasure to acknowledge discussions with S.~P.~Martin and
D.~G.~Robertson, as well as comparisons with their method for addressing the
same problem. The author also thanks S.~Bauberger for sharing details of his
implementation of the dispersion integral method.
This work has been supported in part by the National Science Foundation under
grant no.\ PHY-1519175.


\newpage

\appendix
\section{Divergent parts of master integrals}
\label{sc:div}

\noindent
This section presents analytic results for the divergent part of the master integrals
in Fig.~\ref{fig:diag1}. 
\begin{align}
&U_4(m_1^2,m_2^2,m_3^2,m_4^2) = \frac{1}{\epsilon^3} \sum_{i=2}^{4} \frac{m_i^2}{3} +
\frac{1}{\epsilon^2}\biggl [ -\frac{m_1^2}{6} + \sum_{i=2}^{4} m_i^2
\biggl (\frac{5}{6}-\frac{\log m_i^2}{2} - \frac{\log m_1^2}{2}  \biggr )
\biggr ] \nonumber \\
&\qquad+ \frac{1}{\epsilon} \biggl [
\begin{aligned}[t] &m_1^2\biggl ( -1 + \frac{\log m_1^2}{2} \biggr )
+ \sum_{i=2}^{4} m_i^2 \biggl ( \frac{4}{3}+\frac{\pi^2}{12}-\log m_i^2-
\frac{3}{2}\log m_1^2 \\
&+\frac{1}{4}\log^2 m_i^2 +\frac{1}{4}\log^2 m_1^2
+\log m_i^2 \log m_1^2 \biggr) \biggr ] \end{aligned} \nonumber \\
&\qquad + U_{4,\rm fin}(m_1^2,m_2^2,m_3^2,m_4^2), 
 \displaybreak[0] \\[1ex]
&U_4(0,m_2^2,m_3^2,m_4^2) = -\frac{1}{\epsilon^3} \sum_{i=2}^{4} \frac{m_i^2}{6} +
\frac{1}{\epsilon^2} \sum_{i=2}^{4} m_i^2
\biggl (-\frac{2}{3}+\frac{\log m_i^2}{2}  \biggr )
 \nonumber \\
&\qquad+ \frac{1}{\epsilon} \biggl [
 \sum_{i=2}^{4} m_i^2 \biggl ( \frac{4}{3}+\frac{\pi^2}{24}-\log m_i^2
 +\frac{1}{4}\log^2 m_i^2 \biggr)  - T_{3,\rm fin}(m_2^2,m_3^2,m_4^2) \biggr ] \nonumber \\
&\qquad + U_{4,\rm fin}(0,m_2^2,m_3^2,m_4^2), 
 \displaybreak[0] \\[1ex]
&U_5(m_1^2,m_2^2,m_3^2,m_4^2,m_5^2) = \frac{1}{\epsilon^3} 
\biggl [ \sum_{i=1}^4 \frac{m_i^2}{6} + \frac{m_5^2}{3} \biggr ]
 \nonumber \\
&\qquad + \frac{1}{\epsilon^2} \biggl [ \sum_{i=1}^4 m_i^2 \biggl (
 1-\frac{\log m_i^2}{2} \biggr ) + m_5^2\biggl (\frac{5}{3} - \log m_5^2 \biggr
 ) \biggr ] \nonumber \\
&\qquad+ \frac{1}{\epsilon} \biggl [
\begin{aligned}[t] &\sum_{i=1}^4 m_i^2 \biggl ( \frac{25}{6} + \frac{\pi^2}{24}
- 3\log m_i^2 + \frac{1}{4}\log^2 m_i^2 \biggr ) \nonumber \\
&+ m_5^2\biggl (\frac{17}{3} + \frac{\pi^2}{12} - 5\log m_5^2 
+ \frac{1}{2}\log^2 m_5^2 \biggr ) \\
&+ \frac{1}{2}\Bigl ( (m_1^2+m_2^2-m_5^2) \log m_1^2 \log m_2^2 +
\text{cycl}_{125} \Bigr ) \\[.5ex]
&+ \frac{1}{2}\Bigl ( (m_3^2+m_4^2-m_5^2) \log m_3^2 \log m_4^2 +
\text{cycl}_{345} \Bigr ) \\
&+\lambda_{125}\biggl (\frac{\pi^2}{6} - \frac{1}{2} \log \frac{m_1^2}{m_5^2}
 \log\frac{m_2^2}{m_5^2} + \log u_{125} \log v_{125} - \text{Li}_2\,
  u_{125}
 - \text{Li}_2\, v_{125} \biggr ) \\
&+\lambda_{345}\biggl (\frac{\pi^2}{6} - \frac{1}{2} \log \frac{m_3^2}{m_5^2}
 \log\frac{m_4^2}{m_5^2} + \log u_{345} \log v_{345} - \text{Li}_2\, u_{345}
 - \text{Li}_2\, v_{345} \biggr ) \biggr ] \end{aligned} 
\nonumber \\
&\qquad + U_{5,\rm fin}(m_1^2,m_2^2,m_3^2,m_4^2,m_5^2),
\end{align}
where ``cycl$_{ijk}$'' refers to cyclic permutations of $\{m_i,m_j,m_k\}$, and
\begin{align}
\lambda_{ijk} &= \sqrt{m_i^4+m_j^4+m_k^4-2(m_i^2m_j^2+m_i^2m_k^2+m_j^2m_k^2)}, 
 \label{eq:lijk} \displaybreak[0] \\
u_{ijk} &= \frac{1}{2m_k^2}(m_i^2-m_j^2+m_k^2-\lambda_{ijk}), \\
v_{ijk} &= \frac{1}{2m_k^2}(m_j^2-m_i^2+m_k^2-\lambda_{ijk}).
\end{align}
Furthermore \cite{ibp,Broadhurst:1998rz},
\begin{align}
&U_6(m_1^2,m_2^2,m_3^2,m_4^2,m_5^2,m_6^2) = \frac{1}{\epsilon}2\zeta(3) + 
U_{\rm 6,fin}(m_1^2,m_2^2,m_3^2,m_4^2,m_5^2,m_6^2),
\end{align}
where $\zeta(x)$ is the Riemann zeta function.

For completeness, the divergent terms of the four-propagator
integral $M(1,1,1,1,0,0)$ are also quoted here:
\begin{align}
&M(1,1,1,1,0,0;\,m_1^2,m_2^2,m_3^2,m_4^2) = \nonumber \\
&\qquad \frac{1}{\epsilon^3} 
 \sum_{\substack{i,j=1\\i\neq j}}^{4} \frac{m_i^2 m_j^2}{6} +
\frac{1}{\epsilon^2}\biggl [ -\sum_{i=1}^4 \frac{m_i^4}{12} + 
 \sum_{\substack{i,j=1\\i\neq j}}^{4} m_i^2 m_j^2
 \biggl (\frac{2}{3}-\frac{\log m_i^2}{2}  \biggr )
\biggr ] \nonumber \\
&\qquad+ \frac{1}{\epsilon} \biggl [
\begin{aligned}[t] &\sum_{i=1}^4 m_i^4\biggl ( -\frac{5}{8} + 
 \frac{\log m_i^2}{4} \biggr )
+ \sum_{\substack{i,j=1\\i\neq j}}^{4} m_i^2 m_j^2 
 \biggl ( \frac{5}{3}+\frac{\pi^2}{24}-2\log m_i^2 \\
&+\frac{1}{4}\log^2 m_i^2 
+\frac{1}{2}\log m_i^2 \log m_j^2 \biggr) \biggr ]  + {\cal O}(\epsilon^0) \,.
\end{aligned} 
\end{align}


\section{Analytic results for some master integrals with one and two massive
propagators}
\label{sc:ana}

\noindent
Various special cases of the $U_4$ and $U_5$ integrals with one and two massive
propagators can be computed
analytically to all orders in $\epsilon$ 
using Mellin-Barnes representations. The $\epsilon$ expansions of the
hypergeometric ${}_2F_1$ functions below were performed with the help of the {\tt HypExp}
package \cite{hypexp}, which internally utilizes the {\tt HPL} package
\cite{hpl}.
\begin{align}
U_4(m_1^2,0,0,0) &= (m_1^2)^{1-3\epsilon}\,
e^{3\gamma_{\rm E}\epsilon}\,2\Gamma^2(1-\epsilon)\Gamma(-2+2\epsilon)
 \Gamma(-1+3\epsilon) \\
&= -(m_1^2)^{1-3\epsilon} \biggl [ \frac{1}{6\epsilon^2} + \frac{1}{\epsilon}
 + \frac{100+5\pi^2}{24} \biggr ] + {\cal O}(\epsilon)\,, 
\displaybreak[0]\\[2ex]
U_4(m_1^2,m_2^2,0,0) &= (m_2^2)^{1-3\epsilon}\,e^{3\gamma_{\rm E}\epsilon}\,
\frac{\Gamma(1-\epsilon)\Gamma^2(-1+2\epsilon)\Gamma(-2+3\epsilon)}%
 {\Gamma(-2+4\epsilon)} \nonumber \\
&\qquad \begin{aligned}[b]
\times \biggl [ &\Gamma(\epsilon) \, x^{-\epsilon} \;
 {}_2F_1(\epsilon, -1+2\epsilon, -2+4\epsilon; 1-x) \\
 &- \frac{\Gamma(1+\epsilon)}{2-2\epsilon} \, x^{1-\epsilon} \;
 {}_2F_1(2\epsilon, 1+\epsilon, -1+4\epsilon; 1-x) \biggr ] 
\end{aligned} \label{eq:u4m1m200} \displaybreak[0] \\
&= (m_2^2)^{1-3\epsilon} \biggl [ \frac{1}{3\epsilon^3} 
 + \frac{1}{\epsilon^2} \biggl ( \frac{5-x}{6} - \frac{1}{2} \log x \biggr ) 
 \nonumber \\
&\qquad 
+ \frac{1}{\epsilon} \biggl ( \frac{16+\pi^2}{12}-x+\frac{x-3}{2}\log x +
 \frac{1}{4}\log^2 x \biggr ) \nonumber \\
&\qquad \begin{aligned}[b] + \biggl ( &
 \frac{20+5\pi^2+8\zeta(3)}{24} - \frac{100+\pi^2}{24}x 
 + \frac{72x-84-11\pi^2}{24}\log x \\
 &+ \frac{3-x+4\log(1-x)}{4}\log^2 x - \frac{1}{12}\log^3 x \\
 &+ (x-1+\log x) \,\text{Li}_2(1-x) + 2\,\text{Li}_3(x) \biggr) \biggr ] + {\cal O}(\epsilon)\,,
\end{aligned}
\displaybreak[0]\\[2ex]
U_4(0,0,m_1^2,m_2^2) &= -(m_2^2)^{1-3\epsilon}\,e^{3\gamma_{\rm E}\epsilon}\,
\frac{\Gamma(1-\epsilon)\Gamma(-\epsilon)\Gamma(1+\epsilon)\Gamma^2(2\epsilon)
 \Gamma(-1+3\epsilon)}{\Gamma(2-\epsilon)\Gamma(4\epsilon)} \nonumber \\
&\qquad \times x^{1-\epsilon} \;
 {}_2F_1(1+\epsilon, 2\epsilon, 4\epsilon; 1-x) \\
&= -(m_2^2)^{1-3\epsilon} \biggl [ \frac{1+x}{6\epsilon^3} 
 + \frac{1}{\epsilon^2} \biggl ( \frac{2(1+x)}{3} - \frac{x}{2} \log x \biggr ) 
 \nonumber \\
&\qquad 
+ \frac{1}{\epsilon} \biggl ( \frac{52+\pi^2}{24}(1+x)-2x\log x +
 \frac{x}{4}\log^2 x + (1-x)\,\text{Li}_2(1-x)\biggr ) \nonumber \\
&\qquad \begin{aligned}[b] + \biggl ( &
 \frac{40+\pi^2}{6}(1+x) + \frac{\zeta(3)}{6}(13-11x)
 - \frac{156x-4\pi^2+15\pi^2x}{24}\log x \hspace{-2em} \\
 &+ \frac{2x-(1-3x)\log(1-x)}{2}\log^2 x - \frac{x}{12}\log^3 x \\
 &+ [4(1-x)-(1-2x)\log x]\, \text{Li}_2(1-x) \\
 & -4(1-x)\,\text{Li}_3(1-x) -
 (1-3x)\,\text{Li}_3(x) \biggr)\biggr ] + {\cal O}(\epsilon)\,, \label{eq:u400m1m2}
\end{aligned}
\displaybreak[0]\\[2ex]
U_5(0,0,0,0,m_1^2) &= (m_1^2)^{1-3\epsilon}\,e^{3\gamma_{\rm E}\epsilon}\,
\frac{\Gamma^2(\epsilon)\Gamma^4(1-\epsilon)\Gamma(2-3\epsilon)
\Gamma(-1+3\epsilon)}{\Gamma(2-\epsilon)\Gamma^2(2-2\epsilon)} \\
&= (m_1^2)^{1-3\epsilon} \biggl [ \frac{1}{3\epsilon^3} + \
 \frac{5}{3\epsilon^2} + \frac{1}{\epsilon}\biggl (\frac{17}{3} +
 \frac{5\pi^2}{12} \biggr )
 + \biggl (\frac{49}{3} + \frac{25\pi^2}{12} - \frac{5\zeta(3)}{3} \biggr )\biggr ] 
\nonumber  \\
 &\quad + {\cal O}(\epsilon)\,, 
\displaybreak[0]\\[2ex]
U_5(m_1^2,m_2^2,0,0,0) &= -(m_2^2)^{1-3\epsilon}\,e^{3\gamma_{\rm E}\epsilon}\,
\frac{\Gamma(\epsilon)\Gamma(1-\epsilon)\Gamma^2(2\epsilon)
 \Gamma(-1+3\epsilon)}{(1-\epsilon)(1-2\epsilon)\Gamma(4\epsilon)} \nonumber \\
&\qquad \times x^{1-\epsilon} \;
 {}_2F_1(1+\epsilon, 2\epsilon, 4\epsilon; 1-x) \label{eq:u5m1m2000} 
 \displaybreak[0] \\
&= (m_2^2)^{1-3\epsilon}\biggl [ \frac{1+x}{6\epsilon^3} 
 + \frac{1}{\epsilon^2} \biggl ( 1+x - \frac{x}{2} \log x \biggr ) 
 \nonumber \\
&\qquad 
+ \frac{1}{\epsilon} \biggl ( \frac{100+\pi^2}{24}(1+x)-3x\log x +
 \frac{x}{4}\log^2 x + (1-x)\,\text{Li}_2(1-x)\biggr ) \nonumber \displaybreak[0] \\
&\qquad \begin{aligned}[b] + \biggl ( &
 \frac{60+\pi^2}{4}(1+x) + \frac{\zeta(3)}{6}(13-11x)
 - \frac{300x-4\pi^2+15\pi^2x}{24}\log x \hspace{-2em} \\
 &+ \frac{3x-(1-3x)\log(1-x)}{2}\log^2 x - \frac{x}{12}\log^3 x \\
 &+ [6(1-x)-(1-2x)\log x]\, \text{Li}_2(1-x) \\
 & -4(1-x)\,\text{Li}_3(1-x) -
 (1-3x)\,\text{Li}_3(x) \biggr)\biggr ] + {\cal O}(\epsilon)\,, 
 \label{eq:u5m1m2000e}
\end{aligned}
\intertext{where}
x &= m_1^2/m_2^2.
\end{align}
The expressions for the one-scale integrals can be readily obtained from
available results in the literature, see $e.\,g.$ Ref.~\cite{sv}, while the
$U_4$ integrals with two scales have been studied in
Ref.~\cite{Kalmykov:2006pu}. Similar results as in this section have also been derived independently in
Ref.~\cite{martin}.


\section{Expressions for one- and two-loop integrals}
\label{sc:12}

\noindent
For the reader's convenience, this appendix lists the well-known formulas for
various one- and two-loop functions that are used in this papers. The one-loop
vacuum function is given by
\begin{align}
A_0(m^2) &= \frac{e^{\gamma_{\rm E}\epsilon}}{i\pi^{D/2}} \int d^Dq \;
\frac{1}{q^2-m^2} \\
&= -e^{\gamma_{\rm E}\epsilon} \, (m^2)^{1-\epsilon} \, 
 \Gamma(-1+\epsilon) \\
&= (m^2)^{1-\epsilon} \biggl [ \frac{1}{\epsilon} + 1 + \epsilon\biggl (
 1+\frac{\pi^2}{12} \biggr ) + \epsilon^2\biggl (
 1+\frac{\pi^2}{12}-\frac{\zeta(3)}{3} \biggr ) \biggr ] 
 + {\cal O}(\epsilon^3)\,,
\displaybreak[0]\\[2ex]
B_0(p^2,m_1^2,m_2^2) &= \frac{e^{\gamma_{\rm E}\epsilon}}{i\pi^{D/2}} \int d^Dq \;
\frac{1}{[q^2-m_1^2][(q+p)^2-m_2^2]} \\
&= (p^2)^{-\epsilon} \biggl [ \frac{1}{\epsilon} + 2
 - \frac{\log(rs)}{2} + \frac{r-s}{2}\log\frac{s}{r} \nonumber \\
&\qquad + 
 \lambda(1,r,s) \biggl ( i \pi + \frac{\log(rs)}{2} -
 \log\frac{1-r-s+\lambda(1,r,s)}{2} \biggr ) \biggr ]+ {\cal O}(\epsilon)\,,
\displaybreak[0]\\
B_0(p^2,0,m_1^2) &= (p^2)^{-\epsilon} \biggl [
 \frac{1}{\epsilon} + 2 - r \log r +
 (1-r) \bigl (i\pi - \log (1-r)
 \bigr ) \biggr ] + {\cal O}(\epsilon)\,,
\displaybreak[0]\\[2ex]
B_0(0,m^2,m^2) &= (1-\epsilon)\frac{A_0(m^2)}{m^2}\,, \\
B_0(0,0,0) &= 0\,, 
\intertext{where $r = m_1^2/p^2$, $s= m_2^2/p^2$, and $\lambda(...)$ is given in
eq.~\eqref{eq:lambda}. For a complex number $z = |z| e^{i\varphi}$
the logarithm is defined as}
\log z &= \log|z| + i\varphi, \qquad \varphi \in (-\pi,\pi]\,.
\end{align}
The mass derivative $B_{0,m_1}$ can be expressed in terms of $A_0$
and $B_0$ functions. After expanding in $\epsilon$ one obtains
\begin{align}
B_{0,m_1}(p^2,m_1^2,m_2^2) = \lambda^{-2}(p^2,m_1^2,m_2^2) \biggl [
 &(p^2+m_2^2-m_1^2) \Bigl ( B_{0,\rm fin}(p^2,m_1^2,m_2^2) + \log m_1^2 -2 \Bigr )
 \nonumber \\
 &+ 2 m_2^2 \log \frac{m_1^2}{m_2^2} \biggr ] + {\cal O}(\epsilon)\, .
\end{align}
Finally, the two-loop vacuum integral is given by \cite{2lvac}
\begin{align}
T_3(m_1^2,m_2^2,m_3^2) &= -\frac{e^{2\gamma_{\rm E}\epsilon}}{\pi^D}
 \int d^Dq_1 \, d^Dq_2 \;
 \frac{1}{[q_1^2-m_1^2][q_2^2-m_2^2][(q_1-q_2)^2-m_3^2]} \\
&= e^{2\gamma_{\rm E}\epsilon}\, (m_3^2)^{1-2\epsilon} \, 
 \frac{\Gamma(1+\epsilon)^2}{2(1-\epsilon)(1-2\epsilon)} \biggl [
 \frac{1+x+y}{\epsilon^2} -\frac{2}{\epsilon} \bigl (x \log x + y \log y \bigr )
 \nonumber \\
&\begin{aligned}
\qquad + \biggl (&x\log^2 x + y \log^2 y - (1-x-y)\log x \log y \\
 &+ \lambda(1,x,y) \Bigl ( 2 \log u \log v - \log x \log y -
  2 \text{Li}_2\,u - 2 \text{Li}_2\,v + \frac{\pi^2}{3} \Bigr ) \biggr )
\end{aligned} \nonumber \\
&\begin{aligned}[b]
\qquad - \epsilon\biggl (&\frac{x}{3}\log^3 x + \frac{y}{3} \log^3 y -
\frac{1-x-y}{2}\log x \log y \,\log(xy) \\
 &+ \lambda(1,x,y) \biggl\{ \frac{1}{2}\log x \log y \,\log(xy) 
 + \frac{4}{3} \log^3(1-w) \\
&\quad+ 2\log^2(1-w)\,\bigl (\log(xy) - \log w \bigr ) \\
&\quad+ \log(1-w)\Bigl (\frac{2\pi^2}{3} + \log^2(xy) \Bigr )
 + \frac{4}{3}\log^3 W + \frac{2\pi^2}{3} \log W \\
&\quad -\frac{4}{3}\log^3 U + 2 \log^2 U \,\log \frac{v^2}{y^2} -
\log U \,\Bigl ( \frac{2\pi^2}{3} + \log^2 y \Bigr ) \\
&\quad -\frac{4}{3}\log^3 V + 2 \log^2 V \,\log \frac{u^2}{x^2} -
\log V \,\Bigl ( \frac{2\pi^2}{3} + \log^2 x \Bigr ) \\
&\quad - 2\log x \;\text{Li}_2\frac{u^2}{x}
  - 2\log y \;\text{Li}_2\frac{v^2}{y} + 2 \log(xy) \;\text{Li}_2\,w \\
&\quad + 2\text{Li}_3\frac{u^2}{x} + 2\text{Li}_3\frac{v^2}{y}
 - 2\text{Li}_3\,w - 4\text{Li}_3(1-w) \\
&\quad+ 4\text{Li}_3\,U + 4\text{Li}_3\,V -2 \zeta(3) \biggr\}\biggr)
+ {\cal O}(\epsilon^2) \biggr ]\,,
\end{aligned}
\end{align}
where
\begin{align}
&x = \frac{m_1^2}{m_3^2}, \quad y = \frac{m_2^2}{m_3^2}, \\[1ex]
&u = \tfrac{1}{2}\bigl[1+x-y+\lambda(1,x,y)\bigr], \quad
v = \tfrac{1}{2}\bigl[1-x+y+\lambda(1,x,y)\bigr], \\[1ex]
&w = \Bigl(\frac{u}{x}-1\Bigr)\Bigl(\frac{v}{y}-1\Bigr), \\
&U = \frac{x}{u}(1-w), \quad V = \frac{y}{v}(1-w), \\
&W = \frac{x}{u} + \frac{y}{v} -1.
\end{align}
For $m_1=0$ one obtains the simpler expression
\begin{align}
T_3(0,m_2^2,m_3^2) &= e^{2\gamma_{\rm E}\epsilon}\, (m_3^2)^{1-2\epsilon} \, 
 \frac{\Gamma(1+\epsilon)^2}{2(1-\epsilon)(1-2\epsilon)} \biggl [
 \frac{1+y}{\epsilon^2} -\frac{2}{\epsilon} y \log y
 \nonumber \\
&\begin{aligned}
\qquad + \Bigl (&y \log^2 y + 2(1-y)\,\text{Li}_2(1-y) \Bigr )
\end{aligned} \nonumber \\
&\begin{aligned}[b]
\qquad - \epsilon\Bigl (&\frac{y}{3} \log^3 y + (1-y) \Bigl\{
 \log^2 y \, \log(1-y) - \frac{\pi^2}{3}\log y \\
&\quad+ 2\log y \, \text{Li}_2(1-y)
 + 2\text{Li}_3\,y + 4\text{Li}_3(1-y) - 2\zeta(3) \Bigr\}\Bigr)
\end{aligned} \nonumber \\
&\qquad+ {\cal O}(\epsilon^2) \biggr ]\,.
\end{align}


\end{document}